\documentclass[12pt,nofootinbib]{article}

\usepackage{epsf}
\usepackage{amsmath}
\usepackage{graphicx}
\usepackage{amsfonts}
\usepackage{amssymb}
\usepackage{latexsym}
\usepackage{color}
\usepackage{cite}
\usepackage{comment} 
\usepackage{feynmf}
\input{colordvi.tex}

\newcommand{\be}{\begin{eqnarray}}
\newcommand{\ee}{\end{eqnarray}}
\allowdisplaybreaks[3]
\renewcommand{\thefootnote}{\fnsymbol{footnote}}

\begin{document}

\begin{titlepage}
\begin{flushright}
RESCEU-1/15
\end{flushright}
\begin{center}

\vskip .5in

{\Large \bf
Large tensor mode, field range bound and consistency
in generalized G-inflation
}

\vskip .45in

{\large
Taro Kunimitsu$^{1,2}$\footnote{email:kunimitsu@resceu.s.u-tokyo.ac.jp},
Teruaki Suyama$^1$\footnote{email:suyama@resceu.s.u-tokyo.ac.jp},
Yuki Watanabe$^{1,3}$\footnote{email:watanabe@resceu.s.u-tokyo.ac.jp},\\ 
and Jun'ichi Yokoyama$^{1,2,4}$\footnote{email:yokoyama@resceu.s.u-tokyo.ac.jp}
}

\vskip .45in%

{\em
$^{1}$ Research Center for the Early Universe,  
Graduate School of Science,\\ The University of Tokyo, Tokyo 113-0033, Japan 
}\\
{\em 
$^{2}$ Department of Physics, Graduate School of Science,\\
The University of Tokyo, Tokyo 113-0033, Japan
}\\
{\em
$^{3}$ Department of Physics, National Institute of Technology, Gunma College,\\
 Gunma 371-8530, Japan
}\\
{\em
$^{4}$ Kavli Institute for the Physics and Mathematics of the Universe, 
WPI, UTIAS,\\ The University of Tokyo, Kashiwa, Chiba 277-8568, Japan
}

\end{center}

\vskip .4in

\begin{abstract}
We systematically show that in potential driven generalized G-inflation models, quantum corrections coming from new physics at the strong coupling scale can be avoided, while producing observable tensor modes. The effective action can be approximated by the tree level action, and as a result, these models are internally consistent, despite the fact that we introduced new mass scales below the energy scale of inflation. Although observable tensor modes are produced with sub-strong coupling scale field excursions, this is not an evasion of the Lyth bound, since the models include higher-derivative non-canonical kinetic terms, and effective rescaling of the field would result in super-Planckian field excursions. We argue that the enhanced kinetic term of the inflaton screens the interactions with other fields, keeping the system weakly coupled during inflation.
\end{abstract}

\end{titlepage}

\renewcommand{\thepage}{\arabic{page}}
\setcounter{page}{1}
\setcounter{footnote}{0}
\renewcommand{\thefootnote}{\#\arabic{footnote}}

\section{Introduction}
The Planck 2015 data, combined with the BICEP2/Keck Array data, have placed strong constraints on the level of primordial tensor modes by B-mode polarization of the Cosmic Microwave Background (CMB)~\cite{Ade:2015tva, Ade:2015lrj}, and
upcoming  CMB observations will constrain the amplitude of tensor modes in the CMB at much stronger levels. However, detection of a tensor-to-scalar ratio of $r\lesssim 0.1$ may be still possible in the near future, fixing the energy scale of inflation. 
In the canonical single field inflation models \cite{Yokoyama:2014nga}, 
such a value of $r$ implies that the change in the inflaton value during inflation 
is at least of the order of the (reduced) Planck mass $M_P\equiv (8\pi G)^{-1/2}$ \cite{Lyth:1996im},
\begin{equation}
\Delta \phi \ge 0.6 \left( \frac{\Delta N}{7} \right) {\left( \frac{r}{0.1} \right)}^{1/2} M_P,
\end{equation}
where $\Delta N \simeq 7$ is the number of $e$-folds corresponding to the observable CMB scales~\cite{Planck:2013jfk, Komatsu:2014ioa}.
If we extrapolate this estimation up to the total number of $e$-folds $N \simeq 50-60$,
which is valid in the simple chaotic inflation models~\cite{Linde:1983gd},
the field excursion fairly exceeds the Planck scale. This is called the Lyth bound \cite{Lyth:1996im}.

Super-Planckian field excursions itself may not be a problem, but since new physics is expected to come in at the strong coupling scale of the theory,\footnote{We use the term ``strong coupling scale" in the same sense as the cut-off scale of the effective theory, which may be different scales in general.} we generally expect corrections to the Lagrangian of the form
\begin{equation}\Delta \mathcal{L} \sim \sum_{m=3}^\infty d_m\frac{\phi^{2m}}{\Lambda^{2m-4}},\end{equation}
where $\Lambda$ is the strong coupling scale of the theory. Due to this correction, field excursion beyond $\Lambda$ may hamper long enough inflation or drastically change predictions of the model, leading to the so-called eta problem.
For general relativity and a canonical scalar field with a renormalizable potential, this strong coupling scale lies at $M_P$. 
The original motivation for these corrections was in supergravity~\cite{Lyth:1996im}, but unknown ultraviolet (UV) physics of quantum gravity is usually assumed to generate them~\cite{Baumann:2014nda}. This can be the general situation in other field theories as well, and thus we discuss the issue further in section \ref{1.1}.

If we assume that new physics will arise at the strong coupling scale of the inflaton-graviton sector, it is natural to say that it becomes difficult to realize inflation with the original tree level action beyond that field value. 
This motivates us to consider whether it is possible to 
evade this eta problem
without spoiling the large tensor-to-scalar ratio of ${\cal O}(0.1)$.
This is usually achieved by making $d_m \ll 1$ with introduction of the shift symmetry~\cite{Freese:1990rb, Kawasaki:2000yn} or its generalization~\cite{Burrage:2010cu, Germani:2010hd}, and better with some UV-completion~\cite{Silverstein:2008sg,Marchesano:2014mla,Adams:2006sv,Keltner:2015xda}.\footnote{Chaotic inflation itself provides $d_m \ll 1$ and thus the model can be regarded as technically natural~\cite{Linde:2005ht,Kehagias:2014wza}, but the introduction of new physics at $\Lambda$ may not guarantee the smallness of $d_m$.}

In this paper, we show that under certain assumptions, the generalized G-inflation models \cite{Kobayashi:2011nu}
can be used to achieve $d_m \ll 1$ and thus evade the eta problem,
without spoiling $r={\cal O}(0.1)$.
These models are based on the generalized Galileon theory \cite{Deffayet:2011gz},
which has been shown to be equivalent to the Horndeski theory \cite{Horndeski:1974wa} in \cite{Kobayashi:2011nu},
and they give the most general single scalar theory coupled with gravity whose field equations are of second order.
 
As emphasized in the next sections, we do not attempt to evade the convensional Lyth bound, for which the canonical field with $d_m \sim {\cal O}(1)$ should move sub-Planckian field ranges. Rather, we focus on whether the sub-strong coupling excursion of non-canonical scalar field models are consistent, while $d_m \sim {\cal O}(1)$ for the non-canonical field.
This in turn approximately corresponds to a canonical field with $d_m \ll 1$.

We provide concise expressions [eqs.~\eqref{bound-1} and \eqref{bound-2}] of the necessary condition to realize sub-strong coupling excursion. We then show that for the models we consider, the strong coupling scale lies below but close to the Planck scale, and that the introduction of a new mass scale, $M \ll M_P$, does not introduce significant quantum corrections to the effective action at the inflationary scale.\footnote{See~\cite{Bezrukov:2010jz} for strong coupling scales of the Higgs inflation model that we do not consider in this paper.}
We finally consider several concrete models satisfying the necessary condition and find that sub-strong coupling excursion is indeed realized.

The rest of the paper is organized as follows. In the next section, we clarify the assumptions we make (which are not completely general), and use an explicit example to highlight our points.
In section~\ref{sec:field_excursion}, we calculate the lower bound of the field excursion value in the framework of generalized G-inflation, and specify what is needed to achieve sub-strong coupling field excursion. 
In section~\ref{strongconsistency}, we calculate the strong coupling scales of the models, and show that they lie well above the inflationary energy scale, indicating that the theoretical predictions of generalized G-inflation are robust. We also consider the internal consistency of generalized slow-roll G-inflatinon models, and demonstrate that a Lyth bound-like (or eta) problem does not reappear at the introduced energy scales. 
In section~\ref{specific}, we explicitly calculate the field excursion values for specific models, and show that sub-strong coupling scale field excursions are indeed realized.

As in the example we show below, our study assumes that interactions arise near the strong coupling scale of the fluctuations of the inflaton-gravity sector, which is below but close to the Planck scale as shown in section~\ref{subsec:strong_coupling}. We also assume that quantum field theory can be used up to $\sim M_P$ if the models are internally consistent.

\section{Implicit assumptions, explicit examples}
\label{1.1}
The value of the field excursion has no physical meaning in general, since field values can be scaled arbitrarily. A normalization of the field is needed if we want field excursion values to have any meaning. In this paper, we canonically normalize the field in Minkowski space ({\it i.e.} in a trivial gravitational background), assuming it consistently exists at low energy. The couplings of the inflaton field with other fields are assumed typically to be of $\mathcal{O}(1)$ when this normalization is used.
We also normalize gravity in the same form as general relativity, by utilizing the Einstein-Hilbert action at leading order. The precise form of the normalizations in this paper will be defined in section~\ref{strongconsistency} [eq.~\eqref{eq:normalization}].


The reason we should realize sub-strong coupling field excursions can be seen in the following example. Let us consider the case where an inflaton field $\phi$ exists in a theory in which the strong coupling scale, which is the energy scale above which the unitary description of the low energy theory breaks down and new physics comes in, exists at the scale $\Lambda$. $\Lambda$ should be at least smaller than $M_P$, but is much larger than the inflationary energy scale.
Although we do not know the UV theory beyond the strong coupling scale, the new physics appearing at that scale would typically include new degrees of freedom with mass of $\mathcal{O}(\Lambda)$, which might result in terms of the following form:
\begin{equation}\mathcal{L} = - \frac{1}{2}(\partial \phi)^2-\frac{1}{2}m^2\phi^2 - \sum_{i=1}^n\left(\frac{1}{2}(\partial \chi_i )^2  +\lambda_i \phi^2 \chi_i^2 +\frac{1}{2}c_i\Lambda^2\chi_i^2\right) , \label{lythex}\end{equation}
where $\lambda_i$ and $c_i \ (>0)$ are coefficients of $\mathcal{O}(1)$. 
(We do not claim that this action is a UV completion of any sort, but we will calculate the result of the above Lagrangian for illustrative purposes.)
We want to realize $\phi^2$-inflation in this model, but if we integrate out the $\chi$ field and renormalize the extra contributions, we would obtain a Coleman-Weinberg like contribution to the $\phi$ potential:
\begin{equation}\Delta \mathcal{L} = \sum_{i=1}^n\frac{1}{64\pi^2}\left(c_i\Lambda^2+2\lambda_i \phi^2\right)^2\ln{\left(1 +\frac{2\lambda_i \phi^2}{c_i\Lambda^2}\right)}. \end{equation}
Although we can renormalize the coefficients of the $\phi^2$ and $\phi^4$ terms to the desired value to realize inflation, the potential still receives contributions of the form
\begin{equation}\Delta \mathcal{L} = \sum_{m=3}^\infty d_m\frac{\phi^{2m}}{\Lambda^{2m-4}}\end{equation}
 with $d_m$ of $\mathcal{O}(1)$. The above Coleman-Weinberg term is derived with a constant $\phi$, but in general, derivative couplings suppressed by the same mass scales would appear due to a dynamical $\phi$. These lead to contributions of the form
\begin{equation}\Delta \mathcal{L} = \frac{\tilde{d}_1}{\Lambda^{2}}(\partial \phi)^2\phi^2 + \frac{\tilde{d}_2}{\Lambda^4}(\partial \phi)^4 + \cdots\end{equation}
with $\tilde{d}_m$ of $\mathcal{O}(1)$. 
Since inflation in the model (\ref{lythex}) occurs at super-Planckian field values, these contributions become dominant at field values where inflation should occur, and the predictions of the tree level Lagrangian will be ruined. Note that there is no problem if $\lambda_i \ll 1$ for all the fields that couple to the inflaton so that the contributions are negligible. This is the usual assumption of chaotic inflation~\cite{Linde:2005ht}. 

In order to realize inflation even with $\lambda_i$ of $\mathcal{O}(1)$, the inflaton field values should run ranges much smaller than the strong coupling scale. This situation can be obtained for example by changing the kinetic sector of the Lagrangian. For instance, if we add a Galileon-like interaction~\cite{Deffayet:2010qz, Kobayashi:2010cm} to the inflaton field, the Lagrangian would be of the following form:
\begin{equation}\mathcal{L} =  - \frac{1}{2}(\partial\phi)^2 - \frac{1}{2M^3}(\partial \phi)^2 \Box \phi -\frac{1}{2}m^2\phi^2 -\sum_{i=1}^n \left(\frac{1}{2}(\partial\chi_i)^2+\lambda_i \phi^2 \chi_i^2 +\frac{1}{2}c_i\Lambda^2\chi_i^2\right), \end{equation}
where $M > 0$ is some mass scale. Note that we are still assuming here that the field $\phi$ is normalized so that the coupling $\lambda_i$ is of $\mathcal{O}(1)$.
As we will show later on in sections~\ref{strongconsistency} and ~\ref{specific}, sub-strong coupling field value inflation in this model is possible for the $\phi > 0$ side of the potential if the newly introduced term dominates in the kinetic sector of $\phi$, and thus the model is viable even for $\lambda_i \sim {\cal O}(1)$. The reason this occurs is, due to the enhanced kinetic term of $\phi$, the couplings with $\chi_i$ and the self-coupling (if any) in the potential become effectively small, leading to an effectively flat potential, realizing inflation at small field values. 
The quantum corrections in the potential also become small, and thus the observable spectra of cosmological fluctuations can be estimated from the classical action, {\it e.g.},~\eqref{eq:h3}.
This will be the main point of this paper.

If we approximately canonically normalize the above inflaton field in the quasi-de Sitter background, we can obtain large tensor modes only when the normalized inflaton field moves  a super-Planckian field range, as in the original Lyth bound argument. In this sense, our consideration is by no means an evasion of the Lyth bound (a strong bound on the field range along this line has been derived for general inflation models using the effective field theory of inflation~\cite{Baumann:2011ws}). But physically, the problem of the Lyth bound comes from the $\mathcal{O}(1)$ coupling of the inflaton field with other fields at the strong coupling scale, and this $\mathcal{O}(1)$ coupling does not necessarily occur with the inflaton field canonically normalized during inflation. Rather, we can think of the situation where the other fields couple with the inflaton field canonically normalized in Minkowski space. This is the most crucial assumption we make in this paper.

Since we want sub-strong coupling field inflation, we have to specify the strong coupling scale to judge whether this is possible.
Naively, the strong coupling scale seems to exist at the newly introduced mass scale $M$, and such is the case in Minkowski space. But we will show that during inflation ({\it{i.e.}} in a non-trivial background), the strong coupling scale is at a much higher energy scale, and sub-strong coupling scale inflation is indeed realized.


\section{Field excursion in the generalized G-inflation}\label{sec:field_excursion}
In this section, we calculate the field excursion for slow-roll models in generalized G-inflation \cite{Kobayashi:2011nu}. Generalized G-inflation is the most general set of inflation models that yield field equations with at most second order derivatives.
The action describing the system is given by \cite{Horndeski:1974wa,Deffayet:2011gz,Kobayashi:2011nu}
\begin{equation}
S=\sum_{i=2}^5 \int d^4x~\sqrt{-g}{\cal L}_i,\label{eq:action}
\end{equation}
with
\begin{eqnarray}
{\cal L}_2&=&K(\phi, X),
\\
{\cal L}_3&=&-G_3(\phi, X)\Box \phi,
\\
{\cal L}_4&=&G_4(\phi,X)R
+G_{4X}\left[(\Box\phi)^2-(\nabla_\mu\nabla_\nu\phi)^2\right],
\\
{\cal L}_5&=&G_5(\phi, X)G_{\mu\nu}\nabla^\mu\nabla^\nu\phi
-\frac{1}{6}G_{5X}\left[ (\Box\phi)^3
-3\Box\phi(\nabla_\mu\nabla_\nu\phi)^2
+2(\nabla_\mu\nabla_\nu\phi)^3 \right],
\end{eqnarray}
where $K$ and $G_i$ are arbitrary functions of $\phi$ and $X\equiv-(\partial\phi)^2/2$.
This represents the most general action of gravity and a single scalar field that
yields at most second order field equations for both the scalar and
gravitational fields.

To be specific, in this paper, 
we consider a class of models in which the generalized Galileon functions take the following forms:
\begin{equation}
K(\phi,X)=-V(\phi)+{\cal K}(\phi)X +\frac{1}{2}h_2(\phi)X^2,~~~~~G_i(\phi,X)=g_i(\phi)+h_i(\phi)X.
\label{Lag-exp}
\end{equation}
As we will see later, we will make the higher order interaction terms with $h_i$ dominate over the canonical kinetic term during inflation. There is a priori no reason that even higher order terms in $X$ could not dominate over those, and the form (\ref{Lag-exp}) does not seem at first to be justified. But for non-trivial (non-canonical) dynamics to occur and at the same time be calculable, some power in the interaction terms has to unnaturally dominate the dynamics (say, by a large coefficient) while the rest of the higher order terms are suppressed. We assume here such a situation, and to illustrate the physics, we consider the case where the lowest order non-trivial terms with $h_i$ dominate, and use the form (\ref{Lag-exp}). Extensions for models with higher order interactions follow directly.

Some form of symmetry may be needed to realize such a situation, such as Galileon symmetry \cite{Nicolis:2008in}\footnote{see for example \cite{Pirtskhalava:2015nla} for a recent consideration on the hierarchy among the interaction terms in Galileon models}, but here we will not invoke such specific symmetry and just work under the assumption that one power in the interaction terms dominates over the canonical kinetic term, leading to non-trivial dynamics.

Without loss of generality, we can set $g_3=g_5=0$ since they can be absorbed
by redefinitions of other functions after integration by parts~\cite{Kobayashi:2011nu}. Plus, to get rid of the field redefinition redundancy, we will take
\begin{equation}g_4 = \frac{1}{2}M_P^2,\ \ \mathcal{K}(\phi)=1.\label{eq:normalization}\end{equation}
in later sections, but in this section we will keep $g_4$ and $\mathcal{K}$ explicitly.

We focus on slow-roll inflation in which all the relevant background quantities 
change very little in Hubble time.
The background evolution is described with a flat Friedmann-Lema\^itre-Robertson-Walker  metric.
In particular, we impose the following conditions for $H =\dot{a}/a$ and other functions:
\begin{align}
|{\dot H}| &\ll H^2,~~~~~|{\ddot \phi}| \ll |H{\dot \phi}|, \\
|\dot{\cal K}| &\ll |H{\cal K}|,\quad |\dot{g}_4| \ll |Hg_4|, \quad |\dot{h}_i| \ll |Hh_i|.
\end{align}
Under this approximation, the background equations of motion reduce to \cite{Kobayashi:2011nu}
\begin{eqnarray}
H^2 \simeq \frac{V}{6g_4},~~~~~3HJ \simeq -V_\phi+12 H^2 g_{4\phi},
\end{eqnarray}
where $J$ is given by
\begin{equation}
J \simeq \dot\phi ({\cal K}+X h_2 +3H \dot\phi h_3 +6H^2 h_4+3H^3\dot\phi h_5 ).
\end{equation}
The first slow-roll equation resembles the Friedmann equation.
As is the case in the canonical single-field scenario, inflation is driven by the potential term.
On the other hand, the equation of motion of the inflaton differs from that of the canonical case.
Inflaton dynamics is thus modified and, as we will see later on in the concrete examples, 
friction is enhanced due to self-interaction and gravitational interaction.
The slow-roll equations above allow us to write ${\dot \phi}$ in terms of $\phi$.

\subsection{Field excursion over the CMB scales}
The number of $e$-folds during some period of inflation can be written as
\begin{equation}
N=\int Hdt=\int \frac{H}{\dot \phi} d\phi. 
\end{equation}
Denoting the field excursion during this period by $\Delta \phi$,
we obtain the following inequality,
\begin{equation}
\Delta \phi \ge  N \left( \frac{\dot \phi}{H} \right)_{\rm min}, \label{lyth1}
\end{equation}
where $({\dot \phi}/H)_{\rm min}$ means the minimum of ${\dot \phi}/H$ during the period.
If we consider the period corresponding to the range of CMB scales ($2\le \ell \lesssim 2\times 10^3$),
${\dot \phi}/H$ is almost a constant, which we denote by ${({\dot \phi}/H)}_*$.
In this case, the above inequality becomes an equality,
\begin{equation}
\Delta \phi = N \left( \frac{\dot \phi}{H} \right)_*. \label{delta-phi-CMB}
\end{equation}

In order to relate this with the observed tensor-to-scalar ratio, we now consider an almost flat Friedmann universe with a metric
\begin{align}ds^2 = -{\cal N}^2dt^2 + h_{ij}({\cal N}^idt+dx^i)({\cal N}^jdt+dx^j), \label{eq:adm}
\end{align}
with
\begin{align}
{\cal N} = 1 + \alpha,\quad
{\cal N}_i = \partial_i\beta,\quad
h_{ij} = a^2e^{2\zeta}[e^{\gamma}]_{ij} + 2a^2\partial_i\partial_jE,
\end{align}
where we have chosen the gauge as $\delta\phi=0$ to fix the time diffeomorphism and $E=0$ to fix the spatial diffeomorphism, $\alpha$, $\beta$ and $\zeta$ are scalar perturbations, and $\gamma_{ij}$ is a tensor perturbation satisfying $\nabla^i\gamma_{ij}= h^{ij}\gamma_{ij}=0$.

To second order in the Lagrangian~\eqref{eq:action}, scalar and tensor modes decouple. 
Solving the momentum and Hamiltonian constraint equations, we obtain $\alpha$ and $\beta$ to first order.
Plugging them back in the quadratic Lagrangian, we get after a few integration by parts
\begin{align}
S_{\zeta^2} &= \int d^4x a^3 \left[ {\cal G}_S \dot{\zeta}^2 - \frac{{\cal F}_S}{a^2}(\partial_i\zeta)^2\right], \label{zeta2nd}\\
S_{\gamma^2} &= \int d^4 x a^3 \left[ {\cal G}_T\dot{\gamma}_{ij}^2-\frac{{\cal F}_T}{a^2}(\partial_k\gamma_{ij})^2\right],
\end{align}
where ${\cal F}_S, \cdots$ are defined in \cite{Kobayashi:2011nu}.
In the present slow-roll case, they become
\begin{eqnarray}
&&{\cal F}_S \simeq \frac{X}{H^2} ({\cal K}+h_2X+ 6H^2 h_4)+\frac{4{\dot \phi}X}{H} (h_3+H^2 h_5),~~~{\cal F}_T \simeq 2g_4, \label{eq:calF}\\
&&{\cal G}_S \simeq \frac{X}{H^2} ({\cal K}+3h_2X+6H^2 h_4)+\frac{6{\dot \phi}X}{H} (h_3+H^2 h_5),~~~{\cal G}_T \simeq 2g_4. \label{eq:calG}
\end{eqnarray}
In healthy theories,
all of these must be positive in order to avoid ghosts and gradient instabilities.
It is clear that we should have ${\cal K} > 0$, $h_2 > 0$, $h_4 > 0$, $\dot\phi h_3 > 0$, or $\dot\phi h_5 > 0$ if one of them dominates in ${\cal F}_S$ and ${\cal G}_S$.
For any choice of $h_i$, gravitons are normalized by $2g_4 >0$, which seems to indicate the strong coupling scale of gravity during inflation.
We will stretch this point later on in connection to the quantum gravity scale of generalized G-inflation.

The tensor-to-scalar ratio in the generalized G-inflation models is given by \cite{Kobayashi:2011nu}
\begin{equation}
r= \frac{16 c_s{\cal F}_S}{c_t{\cal F}_T}, \quad
c_s^2 = \frac{{\cal F}_S}{{\cal G}_S}, \quad
c_t^2= \frac{{\cal F}_T}{{\cal G}_T} \simeq 1. \label{G-r}
\end{equation}
Substituting Eqs.~\eqref{eq:calF} and \eqref{eq:calG} into Eq.~(\ref{G-r}) and solving
for ${\dot \phi}$ yields
\begin{equation}
\left|\frac{\dot \phi }{H}\right| = \sqrt{\frac{g_4 r}{8c_s^2}} {\left( {\cal K}+h_2X
+6H^2h_4+4H{\dot \phi} (h_3+H^2 h_5) \right)}^{-1/2},
\label{dot-phi}
\end{equation}
where $c_s^2 \sim {\cal O}(1)$ when one component (or a particular combination of some components) dominates over the others, as can be seen from eq.~(\ref{eq:calF}) - (\ref{G-r}).
Even when some or all of the components are the same order of magnitude, 
$c_s^2$ is generically ${\cal O}(1)$ except for cases of accidental cancellation. Although large powers of $X$ not included in the form (\ref{Lag-exp}) can lead to small $c_s$, the tensor to scalar ratio is proportional to $c_s$ as in (\ref{G-r}), so those models with small $c_s$ lead to unobservable tensor modes, and are not the models of interest here.
In the rest of this paper, we focus our analysis on the models with $c_s^2={\cal O}(1)$ (this is indeed realized for the explicit models in section~\ref{specific}).
Using the relation (\ref{dot-phi}), Eq.~(\ref{delta-phi-CMB}) becomes
\begin{eqnarray}
&&\Delta \phi = \frac{N}{4} {( r^{1/2} q)}_*
\simeq 0.6 \left( \frac{N}{7} \right){\left( \frac{r}{0.1} \right)}^{1/2} q_* , \label{bound-phi1}\\
&&q \equiv {\left\{ \frac{2g_4}{c_s^2 \left[ {\cal K}+h_2X+6H^2h_4+4H{\dot \phi} (h_3+H^2 h_5) \right]} \right\}}^{1/2}, \label{def-q}
\end{eqnarray}
where $q^2 > 0$ is guaranteed by ${\cal F}_S > 0$ and ${\cal F}_T >0$.

For Eistein gravity with a canonical scalar field in which 
${\cal K}=1,~g_4=M_P^2/2$ and all the others being zero except for $V$, $q=M_P$ holds identically.
Our result shows that if $q_*$
is much smaller than $M_P$,
it allows the sub-Planckian field excursion $\Delta \phi \ll M_P$
even for $r={\cal O}(0.1)$. Effectively, this amounts to a rescaling of the field, and in itself has no meaning unless the strong coupling scale is unchanged.\footnote{We thank Masahide Yamaguchi for discussion on this point.} One might wonder if $q$ is the strong coupling scale of (longitudinal) gravitons.
We shall answer this point later in subsection~\ref{subsec:strong_coupling}.

The situation $q \ll M_P$ happens, for instance, for models in which one component
in the denominator of Eq.~(\ref{def-q}) dominates over the others,
assuming the canonical coefficient $g_4=M_P^2/2$.
Changing of $g_4$ only rescales the normalization of gravity, so this normalization is the assumption throughout this paper.
This class includes the G-inflation models ($h_3 \neq 0$) \cite{Kamada:2010qe},
gravitationally enhanced friction (GEF) models ($h_4 \neq 0$) \cite{Germani:2010gm, Germani:2011ua,Germani:2014hqa} and running Einstein inflation models ($h_5 \neq 0$) \cite{Kamada:2012se}.
This consideration indicates that sub-Planckian field excursion, even throughout the total period
of inflation since the CMB scale left the Hubble horizon, is a generic feature 
of the potential-driven generalized G-inflation models.

In section~\ref{specific}, for some concrete models (nonzero constant of $h_2$, $h_3$, $h_4$, or $h_5$), 
we explicitly show that our generic
expectation of the sub-strong coupling scale field excursion with a large tensor-scalar ratio
$r ={\cal O}(0.1)$ is indeed realized. We do not pursue realizing sub-Planckian field 
excursion in the models with small values of $g_4 \ll M_P^2$ at large field values, since those models only amount to a rescaling in the normalization of gravity.

\subsection{Field range bound} 
Let us consider the extension of 
the field excursion to that over the total number of 
$e$-folds $N_*=50-60$ since the CMB scale left the Hubble radius.
If the minimum of $r^{1/2}q$ does not differ significantly from the one evaluated at the CMB scales,
Eq.~(\ref{lyth1}) becomes
\begin{equation}
\Delta \phi \gtrsim \frac{N_*}{4} {( r^{1/2} q)}_*
\simeq 4 \left( \frac{N_*}{50} \right){\left( \frac{r}{0.1} \right)}^{1/2} q_* . \label{bound-1}
\end{equation}

On the other hand, if the minimum of $r^{1/2} q$ differs significantly from the one
at the CMB scale, the above bound becomes invalid.
In this case, we have another bound, which we derive in the following. 
For this purpose, we use the same trick as in \cite{Antusch:2014cpa}.
Without loss of generality, we can set $\phi_{\rm end} < \phi_{\rm min} < \phi_*$, where $\phi_{\mathrm{min}}$ is the field value
that gives the minimum value of $r^{1/2}q$.
We start from the following identity;
\begin{equation}
\frac{d}{d\phi} \left( \frac{\dot \phi}{H} \right)=\frac{\ddot \phi}{H{\dot \phi}}-\frac{\dot H}{H^2}
=\delta+\epsilon,~~~~~\delta = \frac{\ddot \phi}{H{\dot \phi}},~~~\epsilon=-\frac{\dot H}{H^2}.
\end{equation}
We assume the slow-roll conditions $|\delta| \ll 1,~\epsilon \ll 1$.
Integrating the above equation from $\phi_{\rm min}$ to $\phi_*$ yields
\begin{equation}
{\left( \frac{\dot \phi}{H} \right)}_*-{\left( \frac{\dot \phi}{H} \right)}_{\rm min}
=(\phi_*-\phi_{\rm min}) \langle \delta+\epsilon \rangle,
\end{equation}
where $\langle \delta+\epsilon \rangle$ is the average of $\delta +\epsilon$
in the range $\phi_{\rm min} \le \phi \le \phi_*$.
Since we are considering the case in which ${(r^{1/2}q)}_{\rm min}$ is much
smaller than ${(r^{1/2}q)}_*$, 
the second term on the left hand side of the above equation can be safely neglected.
Using the inequality $\Delta \phi \ge \phi_*-\phi_{\rm min}$, we obtain
\begin{equation}
\Delta \phi \ge \frac{1}{4 \langle \delta + \epsilon \rangle} {( r^{1/2} q)}_*
\simeq 8 {\left( \frac{\langle \delta + \epsilon \rangle}{0.01} \right)}^{-1} {\left( \frac{r}{0.1} \right)}^{1/2} q_* .
\label{bound-2}
\end{equation}
Equations~(\ref{bound-1}) and (\ref{bound-2}) show that the lower bound on $\Delta \phi$
is significantly reduced from the Planck mass for $q_* \ll M_P$, which provides the possibility
of realizing sub-Planckian excursion for the total period of inflation $N_* \simeq 60$.

Equation~(\ref{bound-1}) or (\ref{bound-2}) is needed for sub-Planckian field excursion to occur, 
but since they are lower bounds of the field excursion value, they do not assure the realization of sub-Planckian field excursion. 
Hence, we will not directly use the above equations in the following sections, but explicitly calculate the field excursion values
from the Lagrangians in section~\ref{specific}. Note that the models are selected based on the discussion above, focusing on those which can give small values of $q$.

Up to this point, we have considered sub-Planckian field excursions, but what we really need is sub-strong coupling scale field excursions, since that is where new physics would come in. In the next section, we show at which scale the strong coupling scale exists, and show that the models are quantum mechanically consistent up to that scale.

\section{Strong coupling and consistency of the slow-roll G-inflation models}
\label{strongconsistency}

In this section, we calculate the strong coupling scale of the generalized slow-roll G-inflation models, and also calculate the one-loop quantum corrections to see that the models are internally consistent. Since we want to get rid of the redundancy of scaling in the models, we will normalize the models by
\begin{equation}g_4 = \frac{1}{2}M_P^2,\ \ \mathcal{K}(\phi)=1.\label{eq:normalization}\end{equation}
This determines the normalization of both the scalar and tensor degrees of freedom, and the redundancy of scaling is removed. 
We will work with this normalization hereafter.

\subsection{Strong coupling scales}\label{subsec:strong_coupling}
Here, we specify the strong coupling scale, i.e. the energy scale at which perturbative unitarity of scattering processes are violated, in the generalized G-inflation models. The strong coupling scale can in general be read off from the coefficient and number of derivatives in the interaction terms of the Lagrangian (if cancellations were absent), and hence we adopt this method here. We will estimate the strong coupling scale with what we can say from the lowest order interaction terms (cubic terms), and we will sometimes ignore $\mathcal{O}(1)$ coefficients using $\sim$, since we will only be considering leading term effects. A more thorough investigation of the strong coupling scale is under study.

The strong coupling scale for general relativity with a canonical scalar field lies at $M_P$.
Because this fact can be seen most easily by considering the gravitational interactions at third order, we will
first consider the strong coupling scale in generalized G-inflation arising from the tensor three-point interactions.
The cubic Lagrangian for three gravitons was derived in \cite{Gao:2012ib} as
\begin{align}
S_{\gamma^3} =\int d^4 x  a^3\left[ \frac{\dot{\phi}X h_{5}}{12}\dot{\gamma}_{ij}\dot{\gamma}_{jk}\dot{\gamma}_{ki} 
+\frac{{\cal F}_T}{4a^2}\left( \gamma_{ik}\gamma_{jl}-\frac12 \gamma_{ij}\gamma_{kl}\right)\partial_k\partial_l \gamma_{ij}\right],
\end{align}
where the first term possibly provides a lower strong coupling scale compared to the second term with ${\cal F}_T \sim M_P^2$ due to the normalization. When the $h_5$ term dominates in $\mathcal{G}_S$ during inflation, we see from Eq.~(\ref{eq:calG}) that the coefficient becomes
\begin{equation}\dot\phi X h_5 \simeq \frac{\mathcal{G}_S}{6H}\simeq \frac{\epsilon M_P^2}{3H}.\end{equation}
After canonically normalizing $\gamma_{ij}$ to $\widetilde{\gamma}^{(+,\times)} \sim M_P \gamma_{ij}/\sqrt2$ for each helicity mode, we find that the term possibly yields a lower strong coupling scale compared to $M_P$:
\begin{align}\Lambda_{\rm strong\ coupling}^{\rm (tensor)}\sim \sqrt{\frac{M_P H}{\epsilon}}. \label{eq:tensor_strong_h5}\end{align}
This value is larger than the scalar strong coupling which we will see later, so can be ignored when determining the strong coupling scale of the model, but this illustrates the situation in which the strong coupling scale becomes smaller due to the higher order interactions. For the case without the $h_5$ term, the strong coupling scale becomes $\sim M_P$, just as in general relativity with a canonical scalar field.

We next consider the strong coupling scale derived from scalar interactions. For instance, the slow-roll G-inflation model
\begin{equation}\mathcal{L} = \frac{M_P^2R}{2}+X - V +\frac{1}{M^3}X\Box \phi, \label{srG}\end{equation}
which we analyze in detail in section \ref{potg}, has a cubic interaction of the form
\begin{equation}\mathcal{L}_{\delta\phi^3} \sim \frac{1}{M^3}\left(\partial{\delta\phi}\right)^2\Box{\delta\phi}. \label{G3int} \end{equation}
We consider the case where the last term in (\ref{srG}) dominates over the canonical kinetic term of the background field during inflation. Canonically normalizing $\delta\phi$ [see the quadratic action~\eqref{eq:h3_quadratic}] in an almost de Sitter space, which is done by $\widetilde{\delta \phi} \simeq \sqrt{3}D^{\frac{1}{2}}\delta \phi$ with $D=H|\dot{\phi}|/M^3$, we obtain
\begin{equation}\mathcal{L}_{\delta\phi^3} \sim \frac{1}{D^{\frac{3}{2}}M^3}\left(\partial\widetilde{\delta\phi}\right)^2\Box\widetilde{\delta\phi}, \end{equation}
plus subleading terms suppressed by slow-roll.
Using the background slow-roll equation of motion, the denominator of the cubic interaction becomes
\begin{equation}D^{\frac{3}{2}}M^3\simeq \sqrt{\epsilon}M_PH^2, \label{scalarstrong} \end{equation}
and this gives the cubed value of the strong coupling scale, which is much larger than the naive value estimated from the Lagrangian, $M^3$ (see \cite{Kamada:2015sca} for a recent calculation in a relevant model).

At first sight, the above value seems to require further calculation, since the cubic action still includes second order derivatives and might be simplified by integration by parts. Indeed, the cubic scalar interactions in the generalized G-inflation framework have been derived~\cite{Gao:2011qe, DeFelice:2011uc, RenauxPetel:2011sb}, and the formulas in the literature lead to a higher strong coupling scale. The simplest form of the resulting cubic action for $\zeta$ in the unitary gauge ($\delta\phi = 0$) can be found in \cite{DeFelice:2013ar}:
\begin{align}
S_{\zeta^3} &= \int d^4 x a^3 \left[\tilde{\mathcal{C}}_1 M_P^2 \zeta \dot{\zeta}^2 + \frac{1}{a^2}\tilde{\mathcal{C}}_2 M_P^2 \zeta (\partial \zeta)^2 + \tilde{\mathcal{C}}_3 M_P \dot{\zeta}^3
+ {\cal O}(\epsilon^2) \right], \label{cubiczeta}\nonumber\\
\tilde{\mathcal{C}}_1 &= -\frac{3}{c_s^2}\left(\frac{1}{c_s^2}-1\right)\epsilon_s + \frac{6}{c_s^2}\delta\mathcal{C}_7,\quad
\tilde{\mathcal{C}}_2 = -\frac{c_s^2}{3}\tilde{\mathcal{C}}_1,\nonumber\\
\tilde{\mathcal{C}}_3 &= \frac{M_P}{c_s^2 H}\left[\left(\frac{1}{c_s^2}-1 -\frac{2\lambda}{\Sigma}\right)\epsilon_s +4\delta\mathcal{C}_6 -\frac{4}{c_s^2}\delta\mathcal{C}_7\right].
\end{align}
Here, $\epsilon_s$ and $\delta\mathcal{C}_i$ are slow-roll parameters and $\lambda/\Sigma \lesssim \mathcal{O}(1)$, all defined in \cite{DeFelice:2013ar} (and also $\mathcal{F}_1$, $L_1$, $\mu_1$, $w_1$ and $\mathcal{C}_{6,7,8}$ below). 
We see that the $\tilde{\mathcal{C}}_3$ term has an extra factor of $H$ in the denominator, so it would become the leading term in determining the strong coupling scale.
After some calculation and canonical normalization of the curvature perturbation $\zeta$ to $\widetilde{\zeta}\sim \sqrt{\epsilon}M_P\zeta$, we obtain 
\begin{align}
{\cal L}_{\zeta^3} \sim \frac{1}{\sqrt{\epsilon}M_P H}\dot{\widetilde{\zeta}}^3,\label{eq:lag_zeta3}
\end{align}
for $c_s^2 \sim {\cal O}(1)$ ($\epsilon$ here and hereafter should all be a combination of $\epsilon_s$ and other slow-roll parameters such as $\delta \mathcal{C}_i$, but we represent the complex combination with $\epsilon$ for simplicity). This gives a lower bound of the strong coupling scale in slow-roll generalized G-inflation $\sim (\sqrt{\epsilon} M_P H)^{1/2}$, which contradicts with the result of (\ref{scalarstrong}).

But actually, we cannot deduce the strong coupling scale from the above formulas since (\ref{cubiczeta}) was derived by using the linear equation of motion of the fluctuations, or more precisely by a field redefinition of $\zeta$ to eliminate terms proportional to the linear equation of motion, as in the original calculations of the bispectrum\cite{Maldacena:2002vr}. This redefinition produces new terms in the quartic action, and leads to a lower strong coupling scale, which coincides with the estimate obtained in (\ref{scalarstrong}).\footnote{We would like to thank Kohei Kamada for pointing this out.} The cubic action (\ref{cubiczeta}) has additional terms proportional to the linear equation of motion, which starting form equation (3.1) of \cite{DeFelice:2013ar} and after some calculation can be written as
\begin{align}
S_{\zeta^3} \supset& \frac{\delta\mathcal{L}_2}{\delta\zeta}\bigg{|}_1 \widetilde{\mathcal{F}_1} \, , \qquad
\frac{\delta\mathcal{L}_2}{\delta\zeta}\bigg{|}_1\equiv -2M_P^2\left[\frac{d}{dt}\left(a^3 \frac{\epsilon_s}{c_s^2}\dot\zeta\right)-a\epsilon_s \partial^2 \zeta\right],\notag\\
\widetilde{\mathcal{F}_1}\equiv&\  \mathcal{F}_1+\frac{1}{2M_P^2\epsilon_s}\mathcal{C}_6\dot\zeta^2+\frac{1}{M_P^2\epsilon_s}\mathcal{C}_7 \left(\frac{3}{4c_s^2} \dot\zeta^2 +\frac{3}{4a^2}(\partial \zeta)^2 - \frac{3H}{c_s^2} \zeta\dot\zeta\right)  \notag\\
&+ \frac{1}{2M_P^2\epsilon_s}\mathcal{C}_8 \left(\zeta \dot\zeta +\partial_i\zeta \partial^i\partial^{-2}\dot\zeta -\partial^{-2}(\partial_i\dot\zeta\partial^i\zeta+\dot\zeta\partial^2\zeta) \right)+ (\text{higher order in slow-roll}).
\end{align}

In order to erase these terms to obtain the cubic action (\ref{cubiczeta}), we need to (perturbatively) redefine the field as
\begin{equation}
\zeta \rightarrow \zeta - \widetilde{\mathcal{F}_1}.
\end{equation}
This field redefinition erases the cubic terms proportional to the linear equation of motion, but at the same time produces new terms in the quartic action. If we focus on for example the $(\partial \zeta)^2$ terms in $\widetilde{\mathcal{F}_1}$, which for the models we consider is 
\begin{equation} \left(\frac{L_1(L_1\mu_1+12X\dot\phi h_5)}{12 w_1 a^2} + \frac{3\mathcal{C}_7}{4c_s^2\epsilon_s} \right)(\partial \zeta)^2\sim \frac{M_P^2}{H^2}(\partial \zeta)^2, 
\label{dzetadzeta}
\end{equation}
the field redefinition produces a quartic term of the form (plugging in the field redefinition in the quadratic action),
\begin{equation}
\mathcal{L}_{\zeta^4}\sim \frac{\epsilon M_P^2}{H^4}\left(\partial\left(\partial\zeta\right)^2\right)^2 \sim \frac{1}{\epsilon M_P^2H^4}\left(\partial\left(\partial\widetilde{\zeta}\right)^2\right)^2 \label{dim10}
\end{equation}
resulting in the strong coupling scale
\begin{equation}\Lambda_{\rm strong\ coupling}^{\rm(scalar)}\sim (\sqrt{\epsilon}M_P H^2)^{\frac{1}{3}}, \label{eq:scalar_strong}\end{equation}
assuming no cancellation occurs at the quartic order. This matches the estimate (\ref{scalarstrong}). Before the field redefinition, terms that lead to this strong coupling scale existed in the cubic action, and field redefinition merely changed the order of perturbation in which these terms appear. For example, the term (\ref{dzetadzeta}) times the linear equation of motion gives rise to the same strong coupling scale at the cubic order. Thus we can conclude that in general, the strong coupling scale can be calculated from the interaction terms before field redefinition. (Note that in the case of general relativity with a canonical scalar field, cancellations occur at each order of perturbation and should lead to the strong coupling scale $M_P$.)

This does not mean that the cubic action (\ref{cubiczeta}) cannot be used for calculating the bispectrum of the CMB fluctuations.
In the case of the bispectrum, we calculate the expectation value of the three point function, calculated via the three point vertex Feynman diagram. The terms proportional to the linear equation of motion do not contribute to the three point Feynman diagram since all the propagators are on-shell \cite{Arroja:2011yj, Burrage:2011hd}.
On the other hand, here we have in mind scattering processes such as $\zeta\zeta \rightarrow \zeta\zeta$, and terms proportional to the linear equation of motion cannot be ignored since diagrams with internal propagators exist. Since the calculation of the $\zeta\zeta \rightarrow \zeta\zeta$ process requires the quartic order action, and since scattering amplitudes should not change by redefinition of the field anyhow, the strong coupling scale does not change due to the redefinition of the field. 

The above result is the generic case for generalized G-inflation models except for $K(\phi, X)$ models ($G_3=G_4=G_5=0$), which only have first order derivatives in the original action and terms with second order derivatives such as (\ref{G3int}) should not appear.
This fact can explicitly be seen in the flat gauge ($\zeta =0$) action (see equation (55) and (56) of \cite{Arroja:2008ga}), and leads to the strong coupling scale
\begin{equation}\Lambda_{\rm strong\ coupling}^{K(\phi, X)}\sim (\sqrt{\epsilon}M_P H)^{\frac{1}{2}}. \label{eq:k_strong}\end{equation}

For models with second order derivatives, the value is given by (\ref{eq:scalar_strong}). A possible exception of this is the model in which the $h_4$ term dominates, where all the leading terms in (\ref{cubiczeta}) vanish since $c_s^2 = {\cal F}_S/{\cal G}_S \simeq 1$ and $\lambda = \delta \mathcal{C}_6=\delta\mathcal{C}_7=0$, and the dominant cubic terms are suppressed by an extra factor of the slow-roll parameter \cite{Germani:2011ua}. This is due to the proximity of the perturbation structure to the canonical scalar case, although interaction terms that do not exist for the canonical scalar case appear. As stated above, field redefinition introduces new terms in the fourth order action, which apparently gives the scale (\ref{eq:scalar_strong}), but {\it if} cancellation of the leading terms occur just as in the third order action, the quartic action will be suppressed by $\epsilon^2$,
which is the case for general relativity with a canonical scalar \cite{Jarnhus:2007ia}, leading to a higher strong coupling scale,
\begin{align} \Lambda_{\rm strong\ coupling}^{h_4=\mathrm{const.}}\sim (M_P H^2)^{1/3}. \label{eq:h4_strong}\end{align} 
Whether such cancellation occurs (at the quartic order and higher) is of future investigation, but we will assume this value for the calculations of section \ref{GEFex} (This will only affect the numerical values of the section). Note that the above scale was originally estimated from tensor-scalar-scalar interaction term in \cite{Germani:2010gm}.

Similar strong coupling scales as above appear when the tensor-scalar-scalar and tensor-tensor-scalar interaction terms (and their higher-order generalization) are considered, which can be seen using the formulas of \cite{Gao:2012ib}. These interactions terms are at least suppressed by slow-roll factors compared to the scalar interactions, and thus lead to higher strong coupling scales. Hence, we will not go into the details here.

Since we have not studied the derivative structure of quartic and higher-order perturbations, we cannot definitively say that the higher-order perturbations will only allow coefficients that give the same strong coupling scale as the cubic order (or quartic order in the case of $h_4=\mathrm{const.}$). This at least seems to be the case at quartic order in Galileon inflation models \cite{Arroja:2013dya}, but the general case will be of future study. Here, we simply assume that such is the case.

Thus, we conclude this section by stating that the resultant strong coupling scale in the models we consider exist at  \eqref{eq:scalar_strong}, \eqref{eq:k_strong}, and \eqref{eq:h4_strong}.
Perturbative unitarity is not violated as long as this value is much greater than the inflationary energy scale in generalized slow-roll G-inflation.

\subsection{Quantum consistency of the slow-roll G-inflation models}

Since we have introduced a new mass scale $M$ to realize non-trivial dynamics, we need to check whether or not quantum effects change the effective action above this scale, namely at the inflationary energy scale. We need to consider this since the field excursion in these models is at least beyond this newly introduced scale, even if it is smaller than the strong coupling scale.
If quantum corrections become too large, a classical treatment of the models is inconsistent and their predictions cannot be relied upon, or possibly the corrections can even destroy inflation.

We will consider the quantum corrections at one-loop level in the potential driven G-inflation model, pursuing the calculations around a classical field value during inflation. We approximate the metric by a non-dynamical de Sitter background.
In order to accomplish the calculations, we adopt a technique used in \cite{deRham:2014wfa}, which recasts the second order action into the form
\begin{equation}S_2 = \int d^4x \sqrt{-g_{\mathrm{eff}}}\left[-\frac{1}{2}g_{\mathrm{eff}}^{\mu\nu} \partial_\mu \delta \phi \partial_\nu \delta \phi - \frac{1}{2}\widetilde{V}^{\prime\prime}\delta\phi^2\right], \label{effmet}\end{equation}
where the effective metric is written in terms of the background inflaton field and true metric. 
From this action, we can use the heat kernel technique pretending that the effective metric was the true metric, utilizing the covariant perturbation method formulated in \cite{Barvinsky:1990up}. We will then move to comoving gauge and show that the same results hold for  all the slow-roll inflation models that we want to consider.

In reality, we must use the in-in formalism to calculate the quantum corrections, whereas the heat kernel method is formulated in the in-out formalism. 
But for the divergent part of the corrections, the values should be the same in both formalisms, since the divergence of the counterterms should be identical. The finite results in the in-in formalism are obtained by replacing the Feynman Green functions in the calculation of the in-out formalism into retarded Green functions \cite{Barvinsky:1987uw}.

The calculations below are done using a Euclidian-signature metric, transformed by a Wick rotation from the physical metric. The physical results are obtained by rotating back to the Lorentzian-signature metric.

We start by calculating the heat kernel, 
\begin{equation}K(\tau)\equiv \exp{[\tau (\Box_{\rm eff} -\widetilde{V}^{\prime\prime})]},\end{equation}
where $\tau$ is a parameter called the proper time. The trace of the heat kernel expanded in terms of the effective curvature and mass is \cite{Barvinsky:1990up}
\begin{align}
\mathrm{Tr}K(\tau) &= \int \frac{d^4 x \sqrt{g_{\mathrm{eff}}}}{(4\pi\tau^2)^2}\left[1 + \tau\left(\frac{1}{6}R_{\mathrm{eff}} -\widetilde{V}^{\prime\prime}\right)+ \tau^2\left\{\left(\frac{1}{6}R_{\mathrm{eff}} -\widetilde{V}^{\prime\prime}\right)f_1(-\tau\Box_{\mathrm{eff}}) \widetilde{V}^{\prime\prime} \right. \right.\notag \\
&\left.\left.+\widetilde{V}^{\prime\prime}f_2(-\tau\Box_{\mathrm{eff}}) R_{\mathrm{eff}} +R_{\mathrm{eff}} f_3(-\tau\Box_{\mathrm{eff}})R_{\mathrm{eff}} +R_{\mathrm{eff}}^{\mu\nu} f_4(-\tau\Box_{\mathrm{eff}})R_{\mu\nu}^{\mathrm{eff}} \right\}+ \mathcal{O}(R_{\rm eff}^3, \widetilde{V}^{\prime\prime3})\right] \label{trk_original}
\end{align}
where $f_i(-\tau\Box_{\mathrm{eff}})$ are \cite{Mukhanov:2007zz}
\begin{align}
&f_1(\xi) \equiv \int^1_0 e^{-\xi u(1-u)}du, & f_2(\xi) &\equiv -\frac{f_1(\xi)}{6}-\frac{f_1(\xi)-1}{2\xi} \notag \\
&f_3(\xi) \equiv \frac{f_1(\xi)}{32}+\frac{f_1(\xi)-1}{8\xi} - \frac{f_4(\xi)}{8}, & f_4(\xi) &\equiv \frac{f_1(\xi) - 1 + \frac{1}{6}\xi}{\xi^2}.
\end{align}
Expanding in terms of $\tau$, we obtain
\begin{align}
\mathrm{Tr}K(\tau) &= \int \frac{d^4 x \sqrt{g_{\mathrm{eff}}}}{(4\pi\tau^2)^2}\left[1 + \tau\left\{\frac{1}{6}R_{\mathrm{eff}} -\widetilde{V}^{\prime\prime}\right\} \right. \notag \\
&\left.+ \tau^2 \left\{\frac{1}{2}\widetilde{V}^{\prime\prime2} -\frac{1}{6}\widetilde{V}^{\prime\prime} R_{\mathrm{eff}} +\frac{1}{120} R_{\mathrm{eff}}^2 +\frac{1}{60}R_{\mu\nu}^{\mathrm{eff}}R^{\mu\nu}_{\mathrm{eff}}\right\} + \mathcal{O}(\tau^3, R_{\rm eff}^3, \widetilde{V}^{\prime\prime3})\right] \label{trk}
\end{align}
which yields the Schwinger-DeWitt (or Seeley-DeWitt) coefficients.
The correction to the effective action is calculated by
\begin{equation}\Gamma_E[g_{\mu\nu}] =\frac{1}{2}\left.\frac{d}{ds}\right|_{s=0}\left(\frac{1}{\Gamma(s)}\int^\infty_{0}d\tau \tau^{s-1}\mathrm{Tr}K(\tau)\right).\end{equation}
The integral over $\tau$ diverges at the lower end for the first few terms in (\ref{trk}), so we introduce a lower bound $\tau_0$ in the integral, which is the regularization scale, and which we identify with the UV cutoff of the theory. (We can justify the identification of $\tau_0^{-1/2}$ with the cutoff energy $\Lambda_c$ by calculating the divergent part of the effective action using both cutoffs.) The upper end of the integral also seems to diverge due to the higher order $\tau$ terms, but we can show that they converge to give finite results by calclutating from Eq.~(\ref{trk_original}).

 The first and second terms in (\ref{trk}) lead to power-law divergences in terms of $\tau_0^{-1}$ in the effective action. Although the existence of these terms at high energies would be natural, the calculation of these terms from the low energy theory cannot be trusted, since the effective action should not depend on the power of the regularization scale that can be taken arbitrarily \cite{Burgess:1992gx, deRham:2014wfa}. Furthermore, they do not appear in regularization schemes such as dimensional regularization. Hence we will ignore them, under the assumption that these terms do not exist in the complete UV theory.

The log divergent terms on the other hand, cannot be arbitrarily ignored, and resembles the contribution that should come from the UV degrees of freedom \cite{Burgess:1992gx, deRham:2014wfa}. This is due to the mathematical structure of the log divergence, which always keeps the coefficent the same even when changing the regularization scale. We also obtain finite terms which are of the same order as the coefficients of the log divergence \cite{Barvinsky:1990up}, which are physical in any case. Thus, in this paper, we will calculate the log divergent terms as in \cite{deRham:2014wfa}, assuming that they are the leading corrections to the tree level action. Note that the divergent log factor cannot be determined to a pricise value, but we assume that the logarithm of the cutoff scale will give a reasonable estimate~\cite{Nicolis:2004qq}.

The log divergent contribution to the effective action is
\begin{equation}\Gamma \simeq \frac{1}{32\pi^2}\int d^4x \sqrt{g_{\mathrm{eff}}}\left[ \frac{1}{2}\widetilde{V}^{\prime\prime2}- \frac{1}{6}\widetilde{V}^{\prime\prime}R_{\mathrm{eff}} + \frac{1}{120} R^2_{\mathrm{eff}} + \frac{1}{60}R^{\mathrm{eff}}_{\mu\nu}R_{\mathrm{eff}}^{\mu\nu}\right] \ln \frac{ \Lambda_c^2}{\mu^2} \label{notCW1}\end{equation} 
where $\mu$ is the renormalization scale that can be identified with the energy scale under consideration, and $\Lambda_c^2 \sim \tau_0^{-1}$ is the cutoff energy of the theory where new physics comes in, and which we assume is around the strong coupling scale. The divergent log factor should be at most $\mathcal{O}(10)$ (including counter terms if they existed), leading to an order of magnitude prediction of the quantum corrections. This contribution should be subdominant for the inflation models we consider, as we will show in the next subsections.

For inflation models with canonical kinetic terms, we just have
\begin{equation}R_{\rm eff}\simeq 12H^2, \ \ \widetilde{V}^{\prime\prime} = V^{\prime\prime}, \end{equation}
and these contributions can be ignored if
\begin{equation}V^{\prime\prime2}\ll V,\end{equation}
which is satisfied for usual chaotic inflation models without non-renormalizable potential terms, due to the small value of the self-coupling constant.

For the Galileon-type models, the conditions become
\begin{equation}
\sqrt{-g_{\mathrm{eff}}}R_{\mathrm{eff}}^2, \  \sqrt{-g_{\mathrm{eff}}}R^{\mathrm{eff}}_{\mu\nu}R_{\mathrm{eff}}^{\mu\nu}, \ \sqrt{-g_{\mathrm{eff}}}\widetilde{V}^{\prime\prime2}\ll \sqrt{-g}V
\end{equation}
with Lorentzian-signature.
Since the following arguments are of order of magnitude accuracy, we will only show the Lorentzian results in the following subsections.


\subsubsection{Potential driven G-inflation: $h_3={\rm const.}$}
For illustrative purposes, 
we first consider potential driven G-inflation, using the $G_3$ term with $h_3 =-\frac{1}{M^3}$,
\begin{equation}S = \int d^4 x \sqrt{-g}\left[ \frac{M_P^2R}{2}+X + \frac{1}{M^3}X\Box \phi - V(\phi)\right]. \end{equation}
The second order action of the inflaton field is written using the fluctuations around the mean field value during inflation
\begin{equation}\phi = \phi_0(t) + \delta \phi (x, t).\end{equation}
We will ignore the metric perturbations for the moment. This leads to
\begin{equation}S_2 = \int d^4x a^3 \left[\left(\frac{1}{2} - \frac{3H\dot{\phi}}{M^3}\right)\dot{\delta \phi}^2 - \frac{1}{a^2}\left(\frac{1}{2}-\frac{\ddot{\phi}}{M^3} -\frac{2H\dot{\phi}}{M^3}\right)\left(\partial_i \delta \phi\right)^2 - \frac{1}{2}V^{\prime\prime}(\phi) {\delta\phi}^2\right].\label{eq:h3_quadratic}\end{equation}

We define the effective metric $g_{\mu\nu}^{\mathrm{eff}}(\phi, g_{\mu\nu})$ as in \cite{deRham:2014wfa}, so that the second order action can be written in the form

\begin{equation}S_2 = \int d^4x \sqrt{-g_{\mathrm{eff}}}\left[-\frac{1}{2}g_{\mathrm{eff}}^{\mu\nu} \partial_\mu \delta \phi \partial_\nu \delta \phi - \frac{1}{2}\widetilde{V}^{\prime\prime}\delta\phi^2\right]\end{equation}
where the effective metric, curvature, and mass are
\begin{align}
g^{\mathrm{eff}}_{\mu\nu}(\phi_0)& = \mathrm{diag}(A,B,B,B) ,\label{geffABBB}\\
A&=-\left(1 -\frac{2\ddot{\phi}}{M^3} -\frac{4H\dot\phi}{M^3}\right)^{\frac{3}{2}}\left(1-\frac{6H\dot\phi}{M^3}\right)^{-\frac{1}{2}} ,\nonumber\\
B&=a^2\left(1-\frac{2\ddot{\phi}}{M^3} -\frac{4H\dot\phi}{M^3}\right)^{\frac{1}{2}}\left(1-\frac{6H\dot\phi}{M^3}\right)^{\frac{1}{2}} ,\nonumber\\
R^{\rm eff}_{00}&=-\frac{3\ddot{B}}{2B}+\frac{3\dot{B}^2}{4B^2} +\frac{3\dot{A}\dot{B}}{4AB} , \label{R00} \\
R^{\rm eff}_{i0} &=0, \label{Ri0}\\
R^{\rm eff}_{ij}&=\left(\frac{\dot{A}\dot{B}}{4A^2} -\frac{\ddot{B}}{2A} -\frac{\dot{B}^2}{4AB}\right)\delta_{ij}, \label{Rij} \\
R_{\mathrm{eff}}&=-\frac{3\ddot{B}}{AB} + \frac{3\dot{A}\dot{B}}{2A^2B} , \label{Reff}\\
\widetilde{V}^{\prime\prime} &= \frac{\sqrt{-g}}{\sqrt{-g_{\mathrm{eff}}}}V^{\prime\prime}\label{geff} .
\end{align}

Using the heat kernel technique and Schwinger-DeWitt expansion~\cite{Mukhanov:2007zz} to second order in Euclidean signature space,
we obtain the log divergent part of the effective action of the form (\ref{notCW1}).
Using the fact that 
\begin{equation}|H\dot\phi| \gg |\ddot\phi|, M^3\end{equation}
during inflation (the first condition comes from slow-roll, the second comes from demanding the $X\Box \phi$ term to dominate over the canonical kinetic term),
we can calculate the leading contributions to the curvature and effective mass, which are (in Lorentzian signature),
\begin{equation}R_{00}^{\mathrm{eff}}\simeq -3H^2, \quad
R_{ij}^{\mathrm{eff}}\simeq 9a^2H^2\delta_{ij}, \quad
R_{\mathrm{eff}} \simeq \frac{3\sqrt{6}}{D}H^2, \quad
\widetilde{V}^{\prime\prime} \simeq \frac{1}{8\sqrt{3}D^2}V^{\prime\prime} ,\end{equation}
where
\begin{equation}D=\frac{H|\dot{\phi}|}{M^3} \simeq \frac{V^{\prime\frac{1}{2}}}{3M^{\frac{3}{2}}} \gg 1\label{eq:D}\end{equation}
during potential driven G-inflation. The sub-leading terms included in the effective curvatures are suppressed by additional factors of $D$ inverse and slow-roll factors. 

The tree level terms in the action are of order $a^3 M_P^2 H^2$, hence making the contributions from the effective curvatures ($\sim \sqrt{-g_{\mathrm{eff}}}R_{\mathrm{eff}}^2 \sim a^3H^4$) negligible, assuming the log factor to be of $\mathcal{O}(1 - 10)$. 
The contribution from the $\sqrt{-g_{\mathrm{eff}}}\widetilde{V}^{\prime\prime 2}$ term is suppressed by an inverse of $D^2$ compared to $V''$, and even when $V^{\prime\prime2}\sim V$, the term becomes negligible when potential driven G-inflation occurs. 
Note that a small value of the effective mass $V^{\prime\prime2}$ compared to $V$ would also make the contribution negligible, which is assumed in the usual chaotic inflation case.  
Thus we can see that the predictions are unchanged and inflation is realized as long as the Galileon like term completely dominates over the canonical kinetic term. Even though the sub-leading terms include contributions that induce higher order derivatives in the equation of motion, they can be completely ignored during inflation due to the suppression.

If the Lagrangian does not change after inflation, the specific model introduced here would induce ghost/tachyonic instabilities after inflation, but for the moment we assume that some new term in the lagrangian comes to dominate after inflation, as in \cite{Kamada:2013bia}. 

\subsubsection{Calculation using unitary gauge}
The above calculations were done in a non-dynamical gravitational background, but precise calculations should be done in a dynamical spacetime including the metric perturbations. Although the previous section could be seen as an approximation of the calculations in flat gauge ($\zeta = 0, E=0$), we will show that we arrive at the same conclusions in unitary gauge ($\delta\phi = 0, E=0$), this time in a model independent form. 

The curvature perturbation in unitary gauge, $\zeta$, is the only dynamical scalar degree of freedom around the background field values in this gauge. 
The second order action of $\zeta$ is of the form Eq.~(\ref{zeta2nd}), where
\begin{equation}\mathcal{F}_S \simeq c_s^2 \mathcal{G}_S (\sim \epsilon M_P^2)\end{equation}
for slow-roll inflation with $c_s^2=\mathcal{O}(1)$. Transforming the second order action into the form
\begin{equation}S_2 = \int d^4x \sqrt{-g_{\mathrm{eff}}}\left[-\frac{1}{2}g_{\mathrm{eff}}^{\mu\nu} \partial_\mu\widetilde{\zeta} \partial_\nu \widetilde{\zeta}\right],\end{equation}
where $\widetilde{\zeta}\equiv M_P\zeta$, we have
\begin{equation}g_{\mu\nu}^{\mathrm{eff}}=\mathrm{diag}(A,B,B,B) \ ,\ A=-\frac{1}{M_P^2}\sqrt{\frac{\mathcal{F}_S^3}{\mathcal{G}_S}} \ ,\ B=\frac{a^2}{M_P^2}\sqrt{\mathcal{F}_S\mathcal{G}_S} \ .\end{equation}

The one-loop effective action recieves a contribution of the form
\begin{equation}\Gamma \simeq \frac{1}{32\pi^2}\int d^4x \sqrt{-g_{\mathrm{eff}}}\left[\frac{1}{120} R^2_{\mathrm{eff}} + \frac{1}{60}R^{\mathrm{eff}}_{\mu\nu}R_{\mathrm{eff}}^{\mu\nu}\right] \ln \frac{ \Lambda_c^2}{\mu^2}. \label{notCW2}\end{equation} 
Using eqs.~\eqref{geffABBB} - \eqref{geff}, we arrive at
\begin{equation}\sqrt{-g_{\mathrm{eff}}} R_{\mathrm{eff}}^{\mu\nu} R^{\mathrm{eff}}_{\mu\nu} \simeq 36a^3H^4 \frac{1}{c_s^3}  ,\ \sqrt{-g_{\mathrm{eff}}} R_{\mathrm{eff}}^2 \simeq 144a^3H^4 \frac{1}{c_s^3}.\end{equation}
Both of them become contributions of $\sim a^3H^4$ in the effective action, and thus can be ignored when compared to the tree level action. 

The biggest difference between this calculation and the non-dynamical case is that in this gauge, the action is that of a massless scalar, and
the contribution from the mass term is included in the kinetic sector in this gauge. We can explicitly see that they are parametrically suppressed compared to the contribution from the metric.

Note that the models we studied here have no special symmetry to protect the Lagrangian. It has been claimed that the absence of Galileon symmetry in the generalized G-inflation models lead to the generation of terms that do not exist in the original Lagrangian and that those models would be inconsistent \cite{Burrage:2010cu}. Although the first half of the claim is true, we have shown that inconsistency is not always the case, since the Horndeski and non-Horndeski terms newly generated are suppressed by slow-roll parameters, and does not affect the time evolution in this model.

\section{Field excursion and observables in specific models}
\label{specific}
Now that we have shown the consistency of the Generalized slow-roll G-inflation models, we will show that large tensor modes can be generated with sub-strong coupling scale excursions of the field in this framework. We show this by explicitly calculating the field excursion during inflation using specific models. We focus on the potential-driven inflation models that yield $q \ll 1$ in Eq.~\eqref{def-q}. For simplicity, we use a monomial potential
\begin{equation}
V(\phi) = \begin{cases}
\frac{1}{p}m^{4-p}\phi^p&\text{for $p\neq4$}\\
\frac{1}{4}\lambda\phi^4&\text{for $p=4$}
\end{cases}
\end{equation}
and the slow-roll approximation throughout the inflationary phase (note that this might change the field range by an $\mathcal{O}(1)$ factor from the exact results using the full equation of motion, although the conclusions will be the same). For $p=4$, $m^{4-p}$ in the equations should be replaced by $\lambda$. Explicit values are shown using $p=2$ and $N=60$, and we use $M=10^{9} - 10^{11}\mathrm{GeV}$ as pivot scales to show the typical energy scale needed to realize sub-strong coupling scale excursion.
We use the values $\left(\sqrt{\epsilon} M_P H\right)^{1/2}, \left(M_P H^2\right)^{1/3}, \left(\sqrt{\epsilon} M_P H^2\right)^{1/3}$ as the approximate strong coupling scales for $h_2=\mathrm{const.}, h_4=\mathrm{const.}, h_{3,5}=\mathrm{const.}$ respectively when showing the field excursion value, along the arguments of section \ref{subsec:strong_coupling}. Note that, to be precise, we should replace $\epsilon$ by a combination of the slow-roll parameters, but as we will see, $\epsilon$ will only appear with powers smaller than $1$, so the precise value will not matter. We will use $\epsilon$ to show where the slow-roll paramters appear, but the value should be understood as an approximation of the combination of the slow-roll parameters that should have been used. 

The simplest nontrivial example would be to use a polynomial function for $\mathcal{K}(\phi)$ in Eq.~(\ref{Lag-exp}), 
which corresponds to running kinetic inflation \cite{Nakayama:2010kt, Nakayama:2014koa}. 
But as we have stated before, this type of Lagrangian can be transformed into its canonical form by a trivial field redefinition.
Thus, we will start with the case for which the field redefinition into a canonical form is not trivial (although an approximate redefinition into an effectively canonical form during inflation can be done for any model).



\subsection{Potential driven K-inflation: $h_2 = {\rm const}.$}
First, we consider the case where a higher order kinetic term dominates over the canonical kinetic term during inflation (an analysis can be found in \cite{Kamada:2013bia}). We use the lowest order version, which is
\begin{equation}{\cal L} = \frac{M_P^2R}{2}+ X + \frac{1}{2M^4}X^2 - V(\phi). \label{eq:h2}\end{equation}
We assume here that the $X^2$ term dominates over $X$ throughout inflation, which is realized if the condition holds at the end of inflation. The slow-roll equation of motion takes the form
\begin{equation}\dot{\phi} = -\left(\frac{2M^4}{3H}m^{4-p}\phi^{p-1}\right)^{\frac{1}{3}}, \label{eomh2}\end{equation}
while the e-folding number during inflation becomes
\begin{equation}N\simeq\frac{3}{p+4}\left(\frac{m^{4-p}}{6p^2M^4M_p^4}\right)^{\frac{1}{3}} \phi_N^{\frac{p+4}{3}} -\frac{3p}{2(p+4)}, \label{efoldh2}\end{equation}
where $\phi_N$ is the field value $N$ e-folds before the end of inflation. Using the equation of motion (\ref{eomh2}), the Friedmann equation, and the value of $\phi$ calculated from (\ref{efoldh2}), the condition that the $X^2$ term dominates over $X$,
\begin{equation}
X\gg M^4
\end{equation}
at the end of inflation, becomes
\begin{equation}
M_P^{\frac{2p}{p+4}}m^{\frac{2(4-p)}{p+4}}\gg M^{\frac{8}{p+4}}.
\end{equation}

By linear perturbative analysis, the primordial power spectrum is obtained as 
\begin{equation}\mathcal{P}_{\zeta} \simeq \frac{\sqrt{3}M^4}{2\pi^2} \left(\frac{m^{4-p}}{6p^2M^4M_P^4}\right)^{\frac{4}{p+4}}\left(\frac{p+4}{3}N + \frac{p}{2}\right)^{\frac{4(p+1)}{p+4}},\end{equation}
which for $p=2$, $N=60$ and the Planck normalization, $\mathcal{P}_{\zeta}=2.2\times 10^{-9}$, yields
\begin{equation}m\simeq 7.3 \times 10^{-9} \left(\frac{M_P}{M}\right)\left(\frac{\mathcal{P}_\zeta}{2.2\times10^{-9}}\right)^{\frac{3}{4}}M_P .\end{equation}
Substituting this back into $\phi_{N=60}$, we obtain an upper bound for the field excursion
\begin{equation}\Delta \phi \lesssim \phi_{N=60} \simeq 0.24 \left(\frac{M}{10^{11}{\mathrm{GeV}}}\right)\left(\frac{\mathcal{P}_\zeta}{2.2\times10^{-9}}\right)^{-\frac{1}{2}} \left(\sqrt{\epsilon}M_PH\right)^{\frac{1}{2}},\end{equation}
which is sub-strong coupling for $M\lesssim 10^{11}$ GeV. Note that the exact value of $M$ that gives strong coupling scale excursion has no meaning, since effects associated with the strong coupling scale comes in at field values smaller than $\left(\sqrt{\epsilon}M_PH\right)^{1/2}$, as in the example of section~\ref{1.1}.

The scalar spectral index and tensor-to-scalar ratio are written as
\begin{equation}n_s-1=-\frac{8(p+1)}{2(p+4)N+3p}, ~~~r=\frac{16\sqrt{3} p}{2(p+4)N+3p}\end{equation}
which for $p=2$ and $N=60$ are
\begin{equation}n_s=0.967 , ~~~r= 0.076.\end{equation}

\subsection{Potential driven G-inflation: $h_3 = {\rm const.}$}\label{subsec:h3}
\label{potg}
Next, we consider the simplest case of $h_3\neq0$, for which the Lagrangian is of the form
\begin{equation}\mathcal{L} = \frac{M_P^2R}{2}+X +\frac{1}{M^3}X\Box \phi -V. \label{eq:h3}\end{equation}
This corresponds to $h_3 = -1/M^3.$
The dynamics of this class of models was analyzed in \cite{Kamada:2010qe}.\footnote{Their notation is related to ours by $g(\phi)=-h_3(\phi)$. Inflation can take place in the $\dot\phi < 0$ side for $h_3 <0$ since stability of the system requires ${\cal F}_S > 0$ and ${\cal G}_S > 0$. See equations~\eqref{eq:calF} and \eqref{eq:calG}.} 
Although this model needs additional terms during the reheating phase \cite{Kamada:2013bia}, it can be the effective theory during inflation.
Assuming that the $h_3$ term dominates over the canonical kinetic term throughout inflation, 
realized when (using the equation of motion and the field value at the end of inflation)
\begin{equation}\frac{H\dot\phi}{M^3}\gg 1, \ \text{i.e.}\ \ M_P^{\frac{2(p-1)}{p+3}} m^{\frac{2(4-p)}{p+3}} \gg M^{\frac{6}{p+3}},\end{equation}
we have an expression for $\dot{\phi}$ in terms of $\phi$,
\begin{equation}\dot{\phi} = -\sqrt{\frac{M^3 V_\phi}{9H^2}}, \end{equation}
which can be used to calculate the e-folding number during inflation
\begin{align}
N = \int \frac{H}{\dot{\phi}} d\phi
\simeq \frac{1}{M^{\frac{3}{2}}M_P^2}\frac{m^{\frac{4-p}{2}}}{p}\frac{2}{p+3}\phi_N^{\frac{1}{2}(p+3)} -\frac{p}{p+3} .
\end{align}
Thus the field value at $N$ e-folds before the end of inflation is given by
\begin{equation}\phi_N = \left[(p+3)N+p\right]^{\frac{2}{p+3}}\left(\frac{pM^{\frac{3}{2}}M_P^2}{2m^{\frac{4-p}{2}}}\right)^{\frac{2}{p+3}} . \end{equation}

The amplitude of the power spectrum gives one constraint on the parameters, which can be written as
\begin{equation}\left(\frac{\mathcal{P}_{\zeta}}{2.2\times 10^{-9}}\right)^{p+3}= \left(3.5\times 10^6\right)^{p+3}\left(\frac{(p+3)N+p}{2}\right)^{3(p+1)}p^{-6}m^{3(4-p)}M^{3p}M_P^{-12}. \end{equation}
Taking $p=2$, $N=60$, we get
\begin{equation}
m= 3.8 \times 10^{-9} \left(\frac{M}{M_P}\right)^{-1}\left(\frac{\mathcal{P}_\zeta}{2.2\times10^{-9}}\right)^{\frac{5}{6}}M_P.
\end{equation}
Substituting this back into $\phi_{N=60}$, we obtain an upper bound for the field excursion during inflation as
\begin{equation}\Delta\phi \lesssim \phi_{N=60} = 0.20 \left(\frac{M}{10^{10}\mathrm{GeV}}\right)\left(\frac{\mathcal{P}_\zeta}{2.2\times10^{-9}}\right)^{-\frac{2}{3}}\left(\sqrt{\epsilon}M_P H^2\right)^{\frac{1}{3}}\end{equation}
which is smaller than $(\sqrt{\epsilon}M_P H)^{1/2}$ if $M\lesssim 10^{10}$ GeV.

The scalar specral index and tensor-to-scalar ratio depend only on the power of the potential and e-folding number of inflation,
\begin{equation}n_s - 1= -\frac{3(p+1)}{(p+3)N+p}, \ r=\frac{64\sqrt{2}}{3\sqrt{3}}\frac{p}{(p+3)N+p}, \end{equation}
which for the current case is
\begin{equation}n_s= 0.970 \ , \ r=0.11 \ .\end{equation}
Thus, apparent small-field inflation with large $r$ is realized.

\subsection{Gravitationally enhanced frictional (GEF) inflation: $h_4 = {\rm const.}$}
\label{GEFex}
We consider the case $h_4=1/(2M^2)$, which corresponds to
\begin{equation}
 \mathcal{L} = \frac{M_P^2R}{2}+X -V +\frac{1}{2M^2}G^{\mu\nu}\partial_\mu\phi \partial_\nu \phi. \label{eq:h4}
\end{equation}
In this case, the friction term in the scalar field equation is enhanced not by scalar self-interaction but by gravitational interaction. Thus, this class of models is called gravitationally enhanced friction (GEF) models and was analyzed in detail~\cite{Germani:2011ua}.
Assuming that the $h_4$ term dominates over the canonical kinetic term throughout inflation, a condition which after the use of the equation of motion and the formula for the field value can be written as
\begin{equation}H^2\gg M^2, \ \text{i.e.}\ \ M_P^{\frac{2(p-2)}{p+2}}m^{\frac{2(4-p)}{p+2}} \gg M^{\frac{4}{p+2}},\end{equation} 
we have 
\begin{equation}\dot{\phi} =-\frac{M^2 V_\phi}{9H^3}.\end{equation}
From this we can calculate the e-folding number during inflation
\begin{equation}N = \int \frac{H}{\dot\phi}d\phi \simeq -\int\frac{V^2}{M^2M_P^4V_\phi}d\phi = \frac{m^{4-p}}{p^2(p+2)M^2M_P^4}\phi_N^{p+2} -\frac{p}{2(p+2)} .\end{equation}
Solving this for $\phi_N$, we obtain
\begin{equation}\phi_N = \left[\frac{p^2(p+2)M^2M_P^4}{m^{4-p}}\left(N+\frac{p}{2(p+2)}\right)\right]^{\frac{1}{p+2}} .\end{equation}

The primordial power spectrum is written as
\begin{equation}\mathcal{P}_{\zeta} =\frac{m^{2(4-p)}}{12\pi^2p^4M_P^8M^2}\left[\frac{p^2(p+2)M^2M_P^4}{m^{4-p}}\left( N + \frac{p}{2(p+2)}\right)\right]^{\frac{2(p+1)}{p+2}} .\end{equation}
For $p=2$, $N=60$, and $\mathcal{P}_{\zeta}=2.2\times 10^{-9}$, we have a relation between $m$ and $M$,
\begin{equation}m = 3.5\times 10^{-11}\left(\frac{M_P}{M}\right) \left(\frac{\mathcal{P}_\zeta}{2.2\times10^{-9}}\right)M_P.\end{equation}
Substituting this back into $\phi_{N=60}$, we obtain an upper bound for the field excursion
\begin{equation}\Delta\phi\lesssim \phi_{N=60} \simeq 0.68\left(\frac{M}{10^{9}\mathrm{GeV}}\right)\left(\frac{\mathcal{P}_\zeta}{2.2\times10^{-9}}\right)^{-\frac{5}{6}}\left(M_PH^2\right)^{\frac{1}{3}} \end{equation}
If we take $M$ to be much smaller than $10^{9}$ GeV, we have $\Delta \phi \ll (M_PH^2)^{1/3}$ and thus can avoid possible problems at the strong coupling scale field value.

The scalar spectral index and tensor-to-scalar ratio can also be calculated as
\begin{equation}n_s -1 = -\frac{4(p+1)}{2(p+2)N+p}, \ \ \ \ r=\frac{16p}{2(p+2)N+p} , \end{equation}
which for $p=2$ and $N=60$ yields 
\begin{equation}n_s = 0.972 \ , \ \ \ r=0.066 \ .\end{equation}

\subsection{Potential driven $G_5$-inflation: $h_5 = {\rm const.}$}
Finally, we consider the simplest case of $h_5 = -1/M^5$, for which the Lagrangian is of the form
\begin{align}
\mathcal{L} =&\frac{M_P^2R}{2}+ X - V 
- \frac{1}{M^5}XG_{\mu\nu}\nabla^\mu \nabla^\nu \phi \nonumber\\
&+\frac{1}{6M^5} \left[ (\Box\phi)^3 
-3\Box\phi(\nabla_\mu\nabla_\nu\phi)^2
+2(\nabla_\mu\nabla_\nu\phi)^3 \right]. \label{eq:h5}
\end{align}
As far as we know, inflation dynamics of this Lagrangian has not been addressed in the literature (although the model itself was proposed in \cite{Kamada:2012se}). 
Below, we show that inflation with sub-strong coupling excursion is possible in this model.

Assuming that the $h_5$ term dominates over the canonical kinetic term in the equation of motion during inflation, realized when (using the equation of motion)
\begin{equation}H^3\dot\phi \gg M^5, \text{i.e.}\ \ M_P^{\frac{2(2p-3)}{2p+3}} m^{\frac{4(4-p)}{2p+3}} \gg M^{\frac{10}{2p+3}},\end{equation}
we have an expression for $\dot{\phi}$ in terms of $\phi$,
\begin{equation}\dot{\phi} = -\sqrt{\frac{M^5 V_\phi}{9H^4}},\end{equation}
which can be used to calculate the e-folding number during inflation
\begin{align}
N = \int \frac{H}{\dot{\phi}} d\phi
\simeq \frac{2}{\sqrt{3} p^{3/2} (2p+3) } \frac{m^{4-p}}{M_P^3 M^{5/2}}
{\phi_N}^{p+\frac{3}{2}} - \frac{p}{2p+3}. 
\end{align}
Thus the field value $N$ e-folds before the end of inflation is given by
\begin{equation}\phi_N \simeq  {\left[ \sqrt{3} p^{3/2} \left(p+\frac{3}{2}\right) \frac{M_P^3 M^{5/2}}{m^{4-p}} \left(N+\frac{p}{2p+3}\right) \right]}^{\frac{2}{2p+3}}. \end{equation} 

The Planck normalization gives one constraint on the parameters, 
which can be written as
\begin{equation}{\cal P}_\zeta=\frac{1}{16 \pi^2 \sqrt{2} p^{\frac{p+6}{2p+3}}} {\left[ \sqrt{3} \left(p+\frac{3}{2}\right)\left( N + \frac{3}{2p+3}\right) \right]}^{\frac{4p+3}{2p+3}}
{\left( \frac{m}{M_P} \right)}^{\frac{2(p+6)}{2p+3}} {\left( \frac{m}{M} \right)}^{-\frac{5p}{2p+3}}. \end{equation}
For instance, taking $p=2$, $N=60$ yields
\begin{equation}m=2.2\times10^{-12} \left(\frac{M}{M_P}\right)^{-5/3} \left(\frac{\mathcal{P}_\zeta}{2.2\times 10^{-9}}\right)^{\frac{7}{6}}M_P.\end{equation}
Substituting this back into $\phi_{N=60}$, we obtain an upper bound for the field excursion during inflation
\begin{equation}\Delta\phi \lesssim \phi_{N=60} = 4.1\times 10^{-2} {\left(\frac{M}{10^{11}{\mathrm{GeV}}}\right)}^{5/3} \left(\frac{\mathcal{P}_\zeta}{2.2\times 10^{-9}}\right)^{-1}\left(\sqrt{\epsilon}M_P H^2\right)^{\frac{1}{3}},\end{equation}
which is smaller than $(\sqrt{\epsilon}M_P H^2)^{1/3}$ if $M\lesssim 10^{11}$ GeV.

The scalar specral index and tensor-to-scalar ratio depend only on the power of the potential and e-folding number of inflation,
\begin{equation}n_s - 1= -\frac{4p+3}{(2p+3)N + p}, ~~~r=\frac{64\sqrt{2}}{3\sqrt{3}}\frac{p}{(2p+3)N+p} .\end{equation}
which for the current case is
\begin{equation}n_s= 0.974, \ r=0.083.\end{equation}
Thus, apparent small field inflation with relatively large $r$ is realized.


\section{Summary and conclusions}
In this paper, we showed that in the framework of generalized G-inflation, the field excursion of the inflation field can be sub-strong coupling scale, even when the tensor-to-scalar ratio is large enough to be observed. In section \ref{1.1}, we argued that this sort of sub-strong coupling field inflation using non-canonical fields can become important depending on the coupling of the inflaton and other fields.

In section \ref{sec:field_excursion}, we introduced a lower bound formula for the field excursion in generalized G-inflation, and  showed that sub-Planckian, and sub-strong coupling scale excursion would become possible in the Horndeski framework. 

In section \ref{strongconsistency}, we showed that the strong coupling scale of the generalized slow-roll G-inflation models are much higher than the introduced mass scale, as long as $c_s^2 = {\cal O}(1)$. We showed that these models are not destroyed by quantum corrections, despite the naive expectation that the theory will break down at the new energy scale introduced in the classical action. We showed this by calculating the one-loop corrections to the effective action via the heat kernel method, arriving at values much suppressed compared to the tree level action. 
Therefore, we conclude that these models are at least internally consistent, and are worth further investigation. Extending this analysis to kinetic driven models that violate the null energy condition will be of future work.

In section \ref{specific}, we calculated the field excursions for concrete models, namely potential driven K-inflation, potential driven G-inflation, GEF inflation, and running Einstein inflation. In all these models, sub-strong coupling scale field excursions were realized depending on the parameters, while predicting large tensor modes that can be confirmed in the future. The calculations were done in the simplest cases of each type of inflation, and the predictions change for more involved models which have terms higher order in $\phi$ and $X$. Thus, even if the tensor-to-scalar ratio is constrained to smaller values in the future, our analysis will still have significance.
\\

\noindent {\bf Acknowledgments:} 
We thank Masahide Yamaguchi for a useful discussion on rescaling of metric and fields, and Kohei Kamada for discussions on the strong coupling scale. YW also thanks Cristiano Germani for many discussions on strong coupling scales. 
This work is supported by Grant-in-Aid for Scientific Research on Innovative Areas
No.~25103505 (TS) from the Ministry of Education, Culture, Sports, Science and Technology (MEXT), the Program for Leading Graduate Schools, MEXT, Japan (TK), the JSPS Research Fellowship for Young Scientists No.~269337 (YW), the Munich Institute for Astro- and Particle Physics (MIAPP) of the DFG cluster of excellence ``Origin and Structure of the Universe" (YW), and the JSPS Grant-in-Aid for Scientific Research (B) No.~23340058 (JY).

\bibliographystyle{JHEP}
\bibliography{ref}

\end{document}